\documentclass[11pt]{iopart}

\usepackage{graphicx}
\usepackage[utf8]{inputenc}
\usepackage[T1]{fontenc}
\usepackage{mathptmx}
\begin{document}

\title[SiPM proton irradiation for application in cosmic space]{SiPM proton irradiation for application in cosmic space}

\author{S.~Mianowski$^1$, D.M.~Borowicz$^{2,4}$, K.~Brylew$^1$ , A.~Dziedzic$^1$, M.~Grodzicka-Kobylka$^1$,  A.~Korgul$^3$, M.~Krakowiak$^3$, Z.~Mianowska$^1$, A.G.~Molokanov$^2$, M.~Moszynski$^1$,  G.V.~Mytsin$^2$, D.~Rybka$^1$, K.~Shipulin$^2$ and T.~Szczesniak$^1$}

\address{$^1$National Centre for Nuclear Research, Otwock, A. Soltana 7, Poland
}
\address{$^2$Joint Institute for Nuclear Research, Dubna, Joliot-Curie 6, Russian Federation
}
\address{$^3$Faculty of Physics, University of Warsaw,  Warsaw, Pasteura  5, Poland
}
\address{$^4$Department of Medical Physics, Greater Poland Cancer Centre, Poznan, Garbary 15, Poland} 
\ead{s.mianowski@ncbj.gov.pl}

\date{\today}

\begin{abstract}
This paper presents the results of the proton irradiation of silicon photomulipliers (SiPMs) by mono-energetic 170~MeV protons with fluence up to 4.6$\times$10$^{9}$ particles/cm$^2$. In our work, three types of silicon photodetectors from Hamamatsu with areas 3$\times$3~mm$^2$ and different subpixel sizes of 25$\times$25~$\mu$m$^2$, 50$\times$50~$\mu$m$^2$, and 75$\times$75~$\mu$m$^2$ were used. The changes in the SiPM dark count rate (DCR) spectrum before and after irradiation in temperatures in the range of 20~$^\circ$C to -65~$^\circ$C are presented. The influence of the DCR changes on the energy resolution of the 662~keV gamma line from the $^{137}$Cs for a non-irradiated GAGG:Ce (1$\%$) scintillator is investigated. The time period of usability of the SiPM detector irradiated by protons in cosmic space was estimated.

\end{abstract}

\maketitle

\section{Introduction}

\begin{figure}[!hb]
\centering
\includegraphics[width=0.5\textwidth]{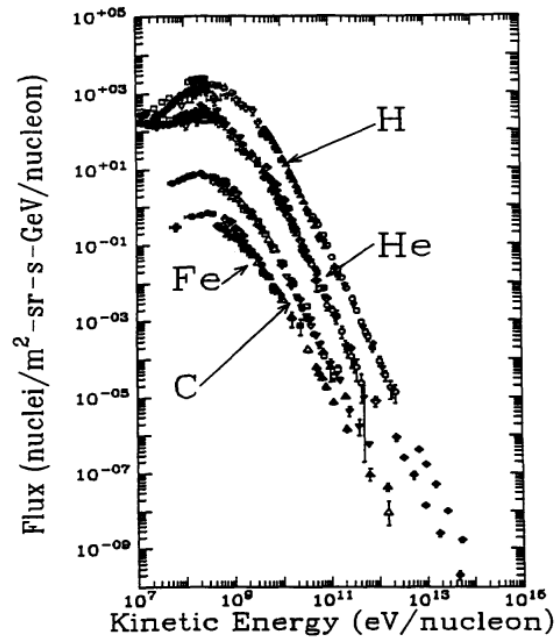}
\caption{\label{fig:cosmics} Measured  cosmic ray energy spectra for the elements H, He, C, and Fe  near the time of solar  minimum  modulation \cite{Simpson}.}
\end{figure}

It is known that about 89$\%$ of nuclear cosmic rays are protons, about 10$\%$ are helium nuclei, and 1$\%$ heavier nuclei \cite{compos}. If we take these particle fluxes as a function of their kinetic energy calculated per single nucleon, as presented in Fig. \ref{fig:cosmics}, it can be found that all nuclei maxima are located in the kinetic energy range of 0.1-1~GeV/nucleon. For protons, 42$\%$ of all of nuclei are located in this energy range and this amount can only increase due to possible shielding. This shows that the lower spectrum of the energy range is the most important for evaluating cosmic radiation damage in photodetectors sent to cosmic space. 

This work is a continuation of our previous research with neutron irradiated silicon photomultipliers (SiPMs) \cite{Mianowski}. The goal of these measurements is to study the changes induced  by the most common particles in cosmic space. Using our data, we try to estimate the usability time of an SiPM irradiated by cosmic protons.

In our work, SiPMs, also called multi-pixel photon counters (MPPCs), from Hamamatsu were used. The changes of dark count rate (DCR) and dark count spectra (DCS) and the degradation of the energy resolution of the spectroscopic gamma line due to proton irradiation are presented. 

\section{Experimental set-up}

Three types of SiPMs with areas 3$\times$3~mm$^2$ and different subpixel sizes were used in our measurements. Table \ref{tab:sipms} presents the basic properties of the chosen SiPMs.

\begin{table}[!b]
\caption{ Properties of Hamamatsu SiPMs used in our measurements \cite{Ham}.}
\label{tab:sipms}
\centering
\begin{tabular}{c|c|c|c|c|c}
\hline
SiPM 			& 	Subpixel				& 	No. of	&	Fill	   &	Ph. det.	 	& Gain  \\ 
type  			&	size      	  			& 	subpixels   &   factor     &    efficiency		&       \\ \hline
S13360-3025CS   &	25$\times$25~$\mu$m$^2$ & 		14400   &     47$\%$   &    25$\%$		& 7.0$\times$10$^5$      \\
S13360-3050CS   &	50$\times$50~$\mu$m$^2$	& 		3600    &     74$\%$   &   	40$\%$		& 1.7$\times$10$^6$      \\ 
S13360-3075CS   &	75$\times$75~$\mu$m$^2$ & 		1600    &     82$\%$   &   	50$\%$		& 4.0$\times$10$^6$      \\ 

\hline
  
\end{tabular}
\end{table}

The irradiation sessions were done at the Dzhelepov Laboratory of Nuclear Problems at the Joint Institute for Nuclear Research (Dubna, Russia). The primary 660~MeV proton beam delivered by the phasotron was moderated to energy of 170~MeV. Flat proton distribution ($\pm$3$\%$) with the size of 8$\times$8~cm$^2$ and a constant flux in the range of 10$^7$ protons/s/cm$^2$ at the beam output was achieved. 

During the irradiation process, the SiPMs were located in the centre of the beam spot, which ensured the same dose for all SiPMs. Each type of SiPM was exposed to proton beam flux with three different irradiation times, which corresponded to three different proton fluences (Tab. \ref{tab:irrtime}) 

\begin{table}[!t]
\caption{ SiPM irradiation time corresponding to obtained proton fluence.}
\label{tab:irrtime}
\centering
\begin{tabular}{c|c|c|c}
\hline
Irradiation & 		Irradiation 		& 		Proton fluence	  		 &   Proton fluence	\\ 
session  	& 		time (s)      		& 		(cm$^{-2}$)              &	 per detector	\\ \hline
1  			& 			30      		& 		3.0$\times$10$^{8}$      &   2.7$\times$10$^{7}$ 				\\
2   		& 			100       		& 		1.0$\times$10$^{9}$      &   9.0$\times$10$^{7}$ 				\\
3    		& 			460 			& 		4.6$\times$10$^{9}$      &   4.1$\times$10$^{8}$ 				\\ \hline
  
\end{tabular}
\end{table}

After the irradiation process, all of the samples were transported to the National Centre for Nuclear Research and were measured in a climatic chamber in the temperature range of 20~$^{\circ}$C to -65~$^{\circ}$C.

\begin{figure}[!hb]
  \centering
  \def\svgwidth{200pt}
  \includegraphics[width=0.7\textwidth]{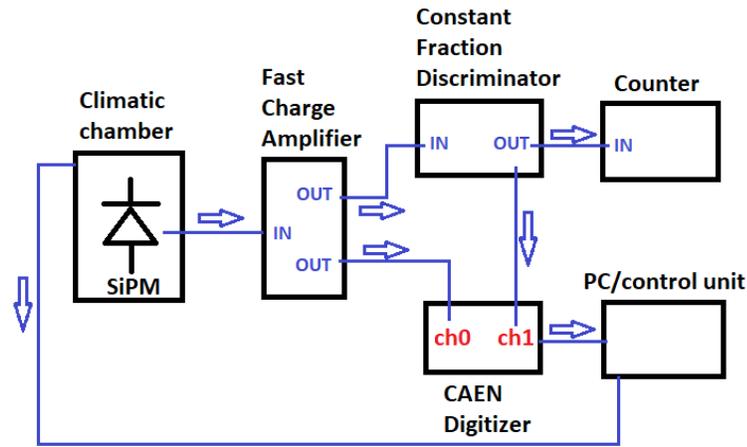}
  \caption{ Scheme of experimental set-up used for DCR and spectrometric measurements. In the second case, a non-irradiated GAGG:Ce scintillator coupled to an SiPM was used.}
  \label{fig:setup}
\end{figure}

Fig. \ref{fig:setup} shows the scheme of the experimental set-up prepared to measure the DCR and DCS at different temperature ranges. The signal from the SiPM was amplified by a transimpedance amplifier and divided by a fast charge amplifier (FCA). The first output from this module was sent to the constant fraction discriminator (CFD), which made it possible to choose signals from the SiPM with single photo-electron (spe) amplitude over a chosen threshold (0.5 spe). Signals that satisfied this condition were counted by the ORTEC Counter. The signals from the second FCA output were sent directly to the 0th channel of the CAEN DT5730 digitizer, which were analyzed using the charge integration (CI) method. For better dead time control in the CAEN DT5730, the CFD logic output signals were also counted by the 1st channel. The disagreement between counts registered by the digitizer and the the ORTEC Counter were in the range of 20$\%$.

\section{Data analysis and results}

\subsection{Dark Count Spectra}

\begin{figure}[!htb]
  \centering
  \def\svgwidth{200pt}
  \includegraphics[width=0.5\textwidth]{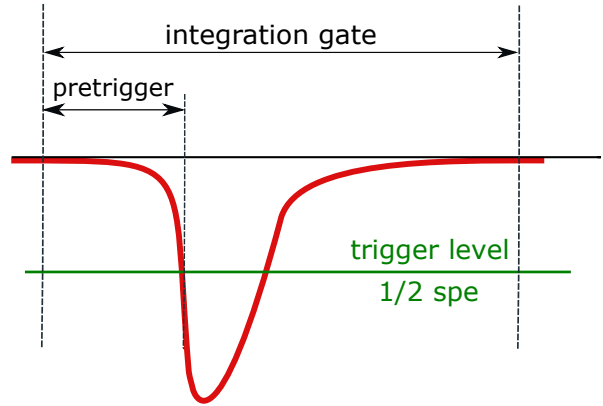}
  \caption{Illustration of the CI method used in our acquisition system.}
  \label{fig:pulse}
\end{figure}

As mentioned above, the CI method was chosen to register DCS. This algorithm, implemented in the CAEN DT5730 digitizer, registered fast SiPM signals without an analog signal shaping modules like the preamplifier or spectroscopic amplifier. The idea of the CI method is presented in Fig. \ref{fig:pulse}.

\begin{figure}[!htb]
  \centering
  \def\svgwidth{200pt}
  \includegraphics[width=0.48\textwidth]{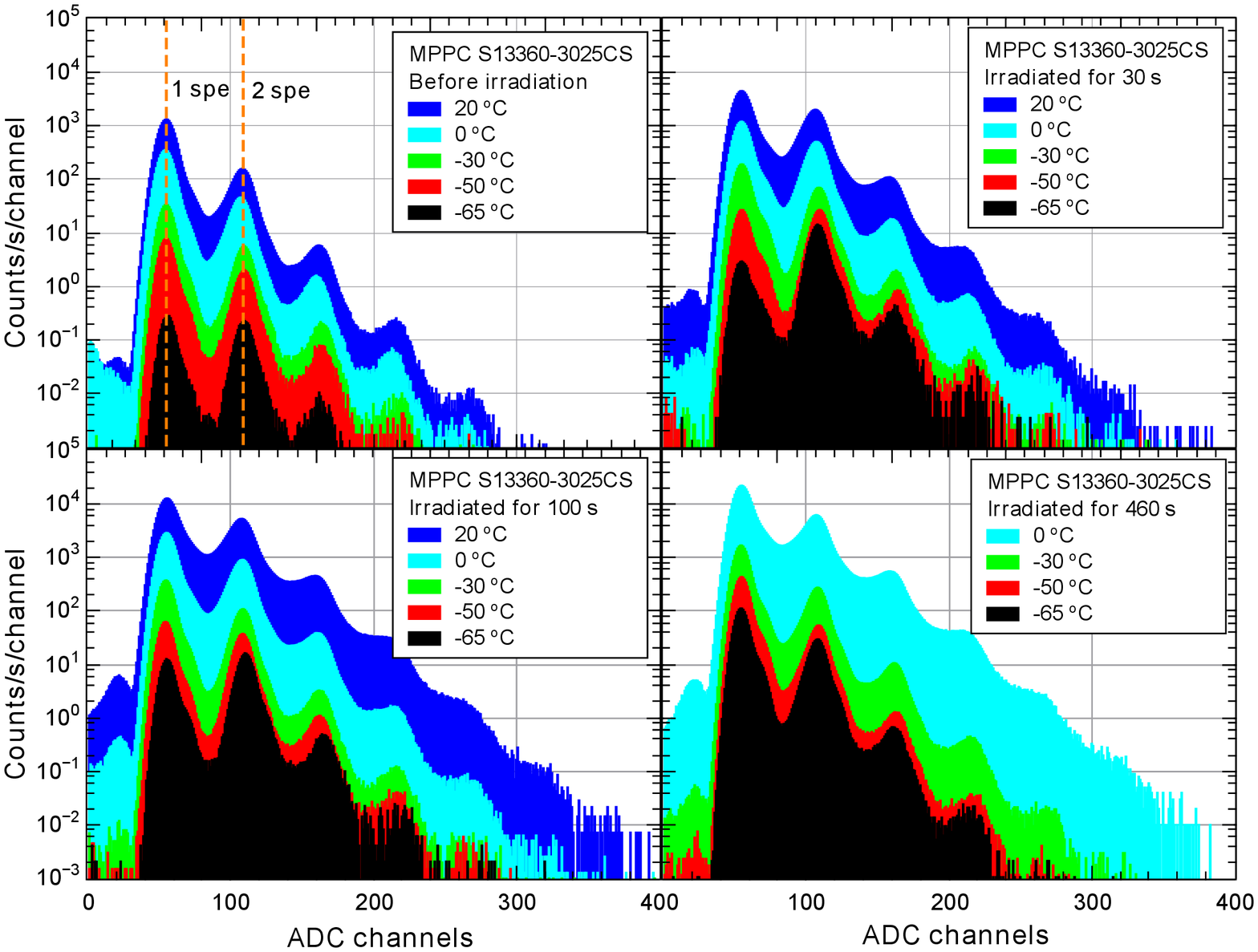}
    \includegraphics[width=0.48\textwidth]{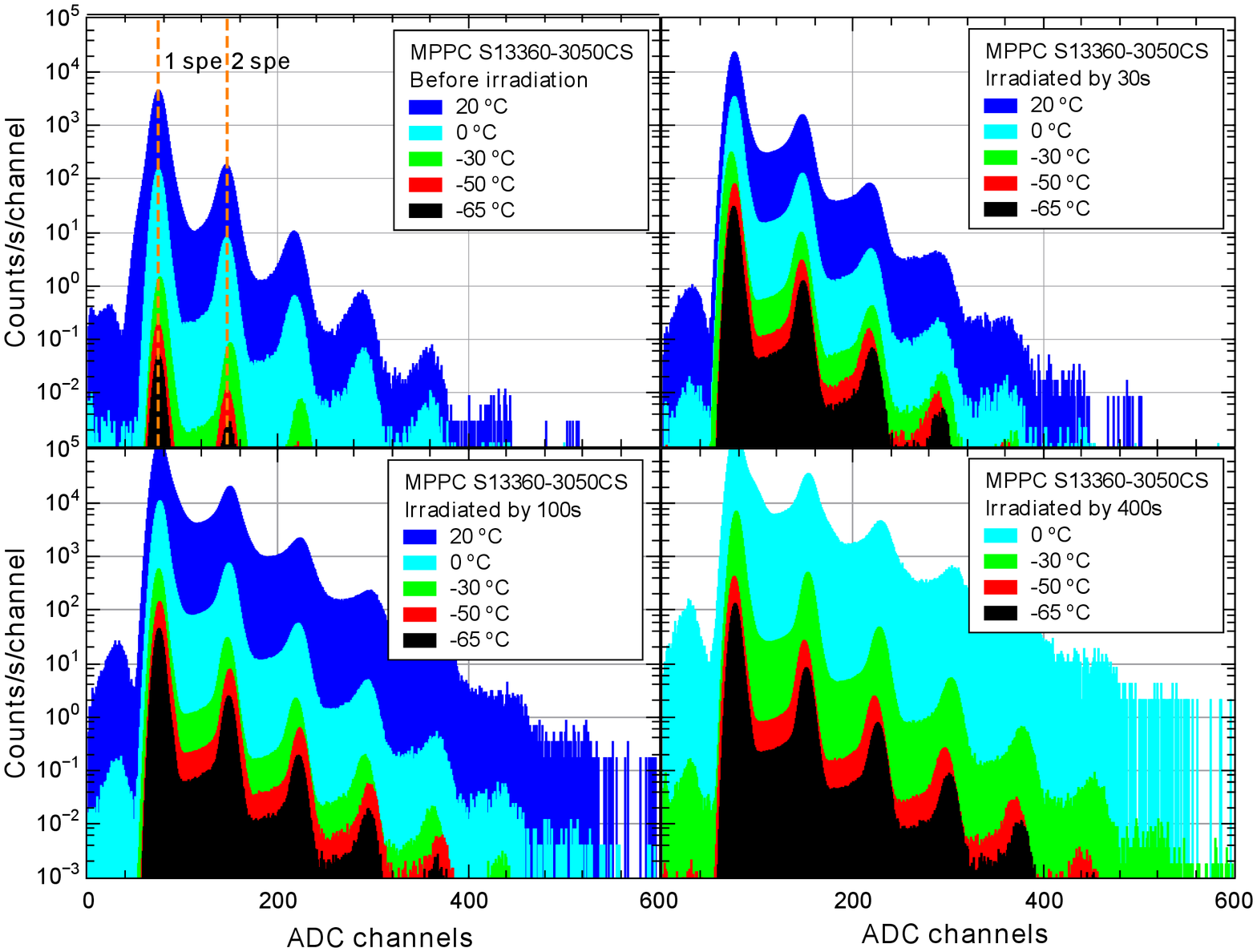}
      \includegraphics[width=0.48\textwidth]{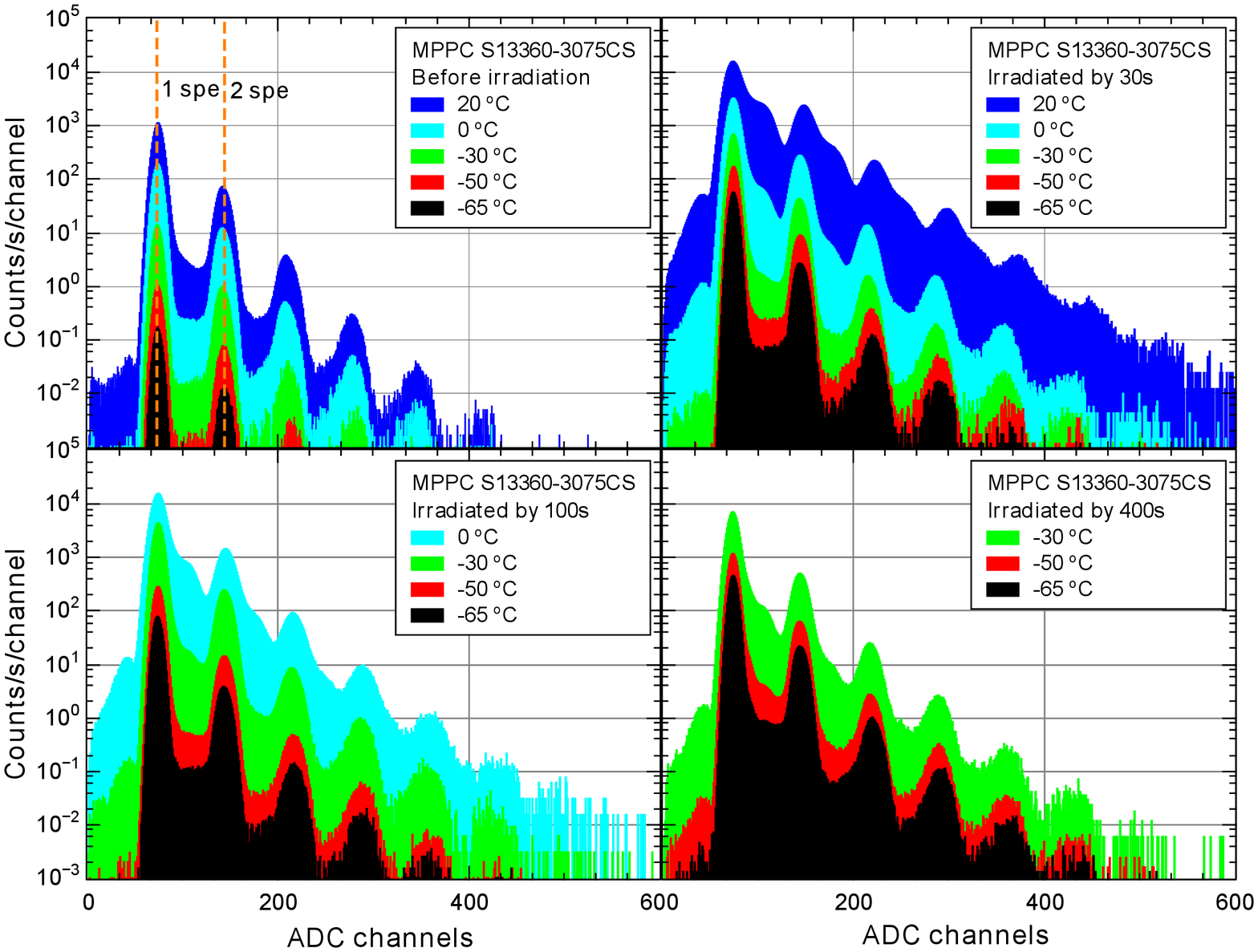}
  \caption{ DCS registered for different proton fluences in various temperatures for the S13360-30XXCS SiPM type with 25$\times$25~$\mu$m$^2$ subpixel size (top), 50$\times$50~$\mu$m$^2$ subpixel size (middle) and 75$\times$75~$\mu$m$^2$ subpixel size (bottom). Vertical dashed lines show the positions of the 1st and 2nd photo-electron peaks.}
  \label{fig:3025-DCR}
\end{figure}

Examples of measured DCS, normalized to counts per second, before and after different irradiation times and for various temperatures are presented in Fig. \ref{fig:3025-DCR}. The same gain for the same SiPM type was kept during all measurements. It was achieved by keeping the same overvoltage (SiPM operating voltage minus breakdown voltage) in various temperatures and is confirmed by the same distance between the 1st and the 2nd photo-electron peak position. In certain cases (irradiated SiPM and temperature of 0~$^\circ$C or 20~$^\circ$C), the digitizer dead time was close to 100$\%$, making it impossible to register the DCS. The DCR was taken from the ORTEC Counter in these cases.  

\begin{figure}[!htb]
  \centering
  \def\svgwidth{200pt}
  \includegraphics[width=0.60\textwidth]{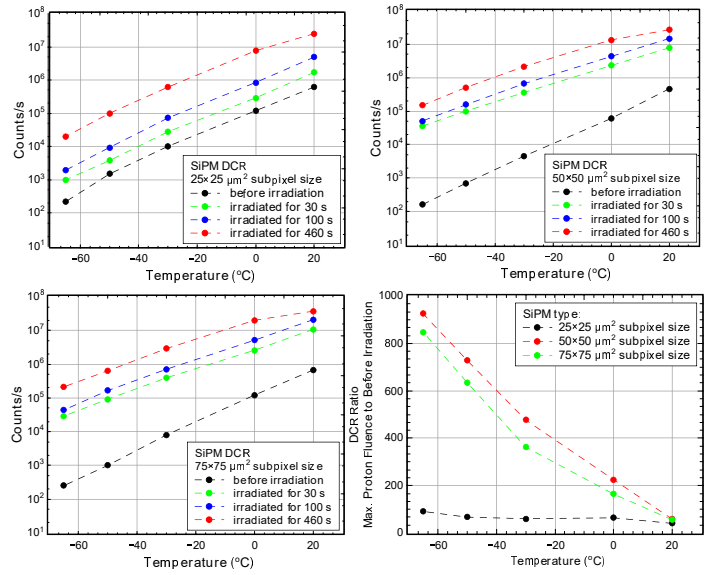}
  \caption{ DCR measured for different SiPM subpixel sizes for different proton fluences in a wide temperature range. Cross-talks and after-pulses are included. Bottom--right - DCR ratio of SiPMs with maximum proton fluence for the non-irradiated case.}
  \label{fig:DCR-all}
\end{figure}

Fig. \ref{fig:DCR-all} shows DCR measured for all SiPM types for different irradiation times in a wide temperature range. As we can see, the lowest DCR change is observed for the smallest subpixel size. On the other hand, the highest DCR reduction with temperature decrease is observed for 50$\times$50~$\mu$m$^2$ and 75$\times$75~$\mu$m$^2$ subpixel sizes. For all cases of irradiated SiPMs, a higher order of photo-electrons (spe$>$3) appeared. This effect is especially important for scintillators coupled to SiPMs, where the relatively low light output is present, like in plastic scintillators, where the light yield is in the range of 10000 photons/MeV. In these cases, DCS and registered radiation events can be seen in the same range of the energy spectrum.

\subsection{Energy resolution}

\begin{figure}[!htb]
  \centering
  \def\svgwidth{200pt}
  \includegraphics[width=0.5\textwidth]{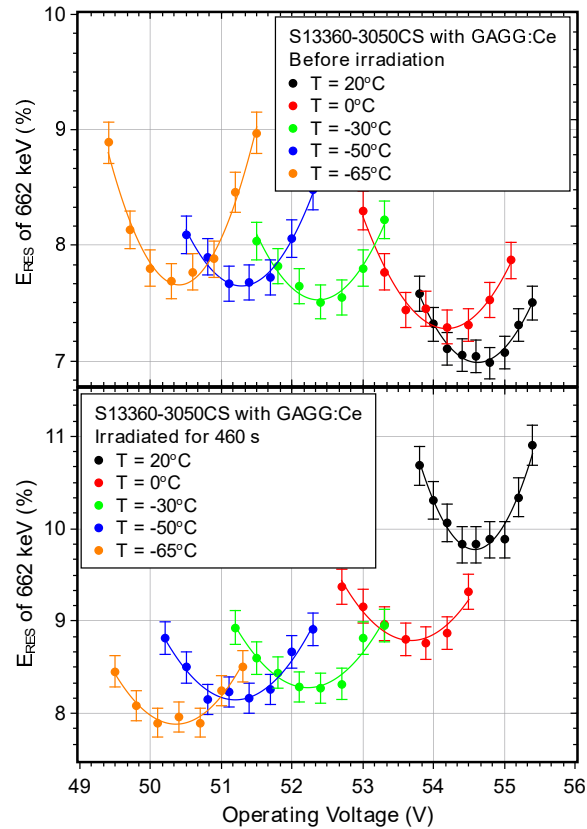}
  \caption{Example of best operating voltage determination, based on the best energy resolution criterion for the S13360-3050CS SiPM in various temperatures.}
  \label{fig:EresVoltage}
\end{figure}

\begin{figure}[!ht]
  \centering
  \def\svgwidth{200pt}
  \includegraphics[width=0.70\textwidth]{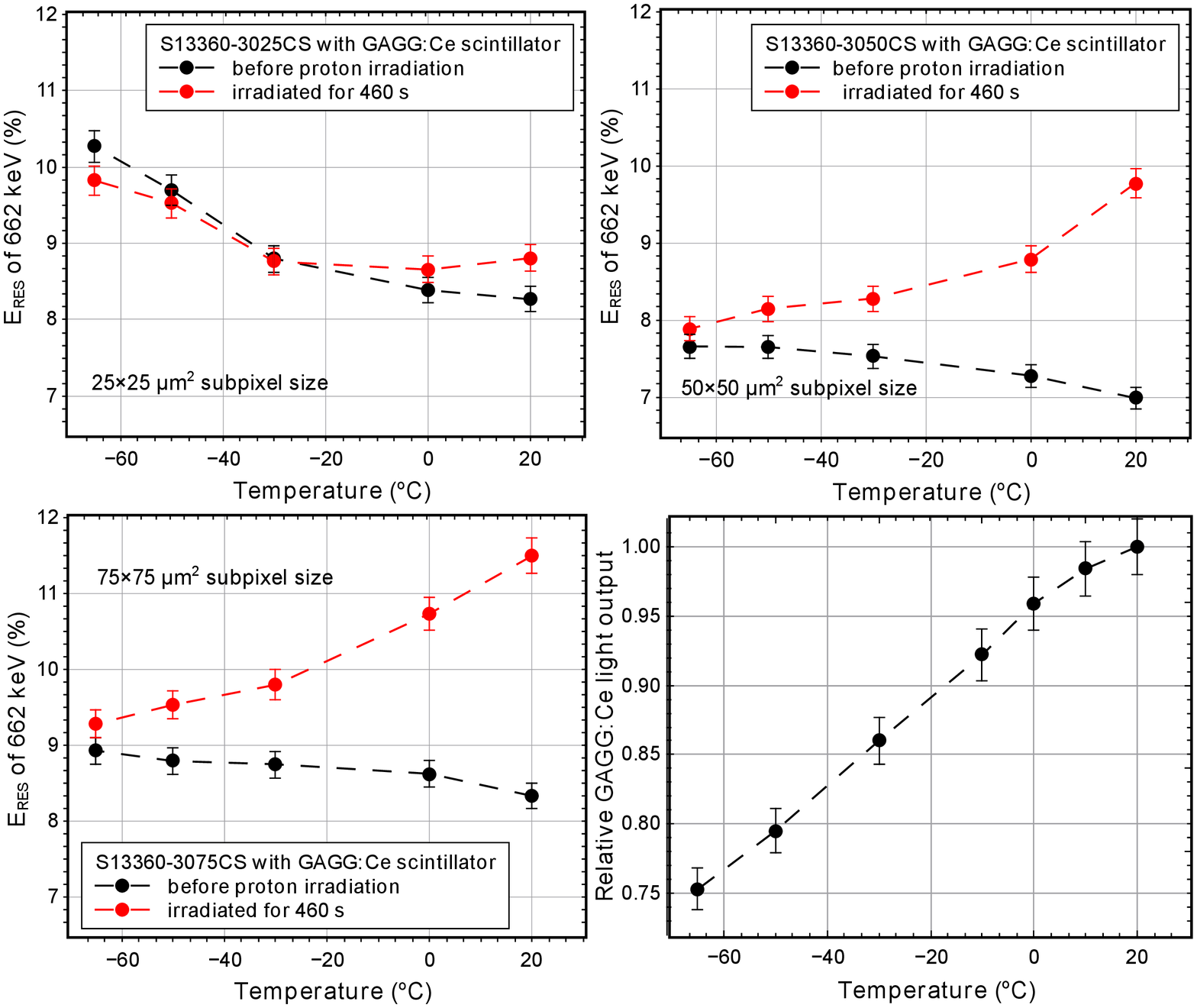}
  \caption{ Energy resolution characteristics for different SiPM subpixel sizes in the temperature range of 20~$^{\circ}$C to -65~$^{\circ}$C. Two cases are shown - SiPMs before irradiation and for the highest proton fluence. Bottom-right -- relative light output of the GAGG:Ce scintillator in the same temperature range. }
  \label{fig:eres}
\end{figure}

The best operating voltage of an SiPM is defined by the value for which the best energy resolution during spectroscopy measurements is obtained \cite{Martyna}. In our case, the 662~keV line from $^{137}$Cs was chosen as a reference point. To obtain the correct value of energy resolution, correction for the non-linearity effect of the SiPM was included. For this reason, three additional gamma lines from $^{22}$Na (511~keV, 1274~keV) and $^{54}$Mn (835~keV) as calibration points were used. 

We determine the energy resolution (E$_{RES}$) of the photodetectors (non-irradiated and irradiated for 460~s) coupled to a non-irradiated GAGG:Ce (1$\%$) scintillator \cite{Miki}. For each SiPM subpixel size, we scanned the operating voltage to find the optimal value. The measurements were performed for SiPMs in the same temperature range as for the DCR determination. An example of E$_{RES}$ determination for an SiPM with a 50$\times$50~$\mu$m$^2$ subpixel size is presented in Fig.~\ref{fig:EresVoltage}.

Obtained energy resolution values are presented in Fig.~\ref{fig:eres}. To understand the energy resolution degradation for the non-irradiated SiPMs, the scintillator light output in this temperature range must be known. This kind of data was measured independently in a climatic chamber with the GAGG:Ce scintillator and a PIN diode. The PIN diode is one of the most suitable photodetectors for this purpose because of its almost flat quantum efficiency in a wide spectral range and low photosensitivity variation to temperature change \cite{Ham, Holl, Mosz1, Mosz2}. Relative light output for the GAGG:Ce is presented in Fig.~\ref{fig:eres} (bottom-right). It is easy to see that the energy resolution degradation for the non-irradiated SiPMs corresponds to the scintillator light output decrease. An analogous situation is observed for all SiPM subpixel sizes.

In the irradiated SiPM case, if we compare the energy resolution change, the 25$\times$25~$\mu$m$^2$ subpixel size is the best choice. We do not see significant differences before and after proton irradiation. The observed characteristics are in agreement if we keep in mind that they correspond to two different units of the same type of SiPM. 

In contrast to the smallest subpixel size, both SiPMs with 50$\times$50~$\mu$m$^2$ and 75$\times$75~$\mu$m$^2$ subpixel sizes show significant energy resolution degradation in the range of 40$\%$. It is worth noting that this situation changes in the low temperature range despite the light output decrease. We can conclude that in the range of -65~$^\circ$C and below, the SiPM energy resolution degradation induced by the DCR increase is significantly reduced.  

\section{Summary}

Our results present the first estimation of SiPM radiation effects induced by protons that can be encountered in cosmic space. We compare SiPM subpixel sizes and corresponding radiation effects, which makes it possible to choose the optimal subpixel size for these kinds of measurements.

The SiPM radiation hardness and energy resolution degradation must be taken into account in gamma spectroscopy measurements whenever experiments are performed in high proton backgrounds. This condition is especially important for scintillators with low light output, where the DCS and registered radiation events can be at the same energy range. We also show that changes induced by radiation can be reduced in a low temperature environment, which is achievable in space conditions.

Using the data presented by Simpson \cite{Simpson} we estimated the total flux of 0.1-1 GeV/nucleon protons incoming from a 4$\pi$ angle. This resulted in a value of  $\sim$2~protons/s/cm$^{2}$, which corresponds to $\sim$0.18~protons/s/detector. Knowing the value of the highest proton fluence per detector obtained in our experiment (4.1$\times$10$^8$), and assuming similar radiation effects for the chosen kinetic energy range \cite{NIEL} \footnote{The same range of the hardness factor in the non-ionizing energy loss scaling hypothesis.} and temperature of -65~$^\circ$C (no significant degradation), the estimated time period of usability for this SiPM detector in cosmic space near the time of solar minimum modulation (larger galactic cosmic ray fluxes) is in the range of 10$^9$~s, which results in a time scale of years. The heavier cosmic ray components are not included in this calculation. Their influence on SiPM radiation hardness is the target of our future experiment.

\ack{This work was supported by Research Program for the Research Group at JINR and Research Centers in Poland: 04-2-1132-2017/2019.
}

\section*{References} 
\bibliography{iopart-num}

\providecommand{\newblock}{}
\begin{thebibliography}{10}
\expandafter\ifx\csname url\endcsname\relax
  \def\url#1{{\tt #1}}\fi
\expandafter\ifx\csname urlprefix\endcsname\relax\def\urlprefix{URL }\fi
\providecommand{\eprint}[2][]{\url{#2}}

\bibitem{compos}
Rigden J~S 1996 {\em Macmillan Encyclopedia of Physics\/} (Simon and Schuster
  Macmillan)

\bibitem{Mianowski}
Mianowski S, Baszak J, Gledenov Y~M, Kopatch Y~N, Mianowska Z, Moszynski M,
  Sibczynski P and Szczesniak T 2018 {\em Nuc. Instrum. Methods Phys. Res. A\/}
  {\bf 906} 30--36

\bibitem{Simpson}
Simpson J~A 1983 {\em Composition and Origin of Cosmic Rays\/} (D. Reidel
  Publishing Company)

\bibitem{Ham}
Hamamatsu  {w}ww.hamamatsu.com

\bibitem{Martyna}
Grodzicka M, Moszynski M, Szczesniak T, Szawlowski M, Wolski D and Baszak J
  2012 {\em IEEE Trans. Nucl. Sci.\/} {\bf 59} 3294–3303

\bibitem{Miki}
Sibczynski P, Iwanowska-Hanke J, Moszynski M, Swiderski L, Szawlowski M,
  Grodzicka M, Szczesniak T, Kamada K and Yoshikawa A 2015 {\em Nuc. Instrum.
  Methods Phys. Res. A\/} {\bf 772} 112--117

\bibitem{Holl}
Holl I, Lorenz E and Mageras G 1988 {\em IEEE Trans. Nucl. Sci.\/} {\bf 35}
  105--109

\bibitem{Mosz1}
Moszynki M, Kapusta M, Mayhugh M, Wolski D and Flyckt S 1997 {\em IEEE Trans.
  Nucl. Sci.\/} {\bf 44} 1052--1060

\bibitem{Mosz2}
Moszynski M, Szczesniak T, Kapusta M, Szawlowski M, Iwanowska J, Gierlik M,
  Syntfeld-Kazuch A, Swiderski L, Melcher C~L, Eriksson L~A and Glodo J 2010
  {\em IEEE Trans. Nucl. Sci.\/} {\bf 57} 2886--2896

\bibitem{NIEL}
Huhtinen M and Aarnio P 1993 {\em Nuc. Instrum. Methods Phys. Res. A\/} {\bf
  335} 580

\end{thebibliography}

\end{document}